\documentstyle[aps,prl,psfig,epsfig,floats,twocolumn,tabularx]{revtex}

\begin{document}
\wideabs{
\title{The Field-Tuned Superconductor-Insulator Transition with and without Current Bias}

\draft


\author{E. Bielejec and Wenhao Wu}
\address{Department of Physics and Astronomy, University of Rochester,}
\address{Rochester, New York 14627}

\date{\today}

\maketitle

\begin{abstract}

The magnetic-field-tuned superconductor-insulator transition has
been studied in ultrathin Beryllium films quench-condensed near 20
K. In the zero-current limit, a finite-size scaling analysis
yields the scaling exponent product $\nu z$ $=$ 1.35 $\pm$ 0.10
and a critical sheet resistance $R_{c}$ of about 1.2$R_{Q}$, with
$R_{Q} = h/4e^{2}$. However, in the presence of dc bias currents
that are smaller than the zero-field critical currents, $\nu z$
becomes 0.75 $\pm$ 0.10. This new set of exponents suggests that
the field-tuned transitions with and without dc bias currents
belong to different universality classes.

\end{abstract}

 }

%\begin{multicols}{2}

The superconductor-insulator (SI) transition in ultrathin films is
believed to be a continuous quantum phase transition \cite{Sondhi}
occurring at $T =$ 0 as the quantum ground state of the system is
tuned by varying disorder, film thickness, magnetic field, or
carrier concentration through a critical value. Models of the SI
transition can be grouped into two categories: ones that rely on
fluctuations in the phase \cite{Fisher,Fisher2,Fisher3,Cha} and
others that rely on fluctuations in the amplitude \cite{Valles} of
the order parameter to drive the transition. For models in the
first category, Cooper pairs are assumed to be present on the
superconducting side as well as on the insulating side of the
transition. Based on this "dirty-boson" model, Fisher $\em{et}$
$\em{al.}$ \cite{Fisher2,Fisher3} have proposed a phase diagram
for a two-dimensional system as a function of disorder and
magnetic field, with the transition being described in terms of
interacting bosons (the Cooper pairs) in the presence of disorder.
In the vicinity of the quantum critical point, the resistance is
predicted to obey the scaling law \cite{Fisher2,Fisher3}:
\begin{equation}
R(\delta, T) \propto R_{c} f(\delta t),
\end{equation}
where $t(T)$ $=$ $T^{-1/\nu z}$, and $\delta$ is the deviation of
a tuning parameter from its critical value. For field-tuned
transitions, $\delta = \vert B - B_{c} \vert$  with $B$ and
$B_{c}$ being the applied magnetic field and the critical field,
respectively. The critical resistance $R_{c}$ is predicted to have
a universal value of $R_{Q} = h/4e^{2} \approx$ 6.5
k$\Omega/\Box$. Scaling arguments set a lower bound on the
correlation length exponent $\nu$ $\ge$ 1 \cite{Sondhi} and give
the value of the dynamical critical exponent $z$ $=$ 1
\cite{Fisher2,Fisher3}. Experimentally, critical exponents
consistent with these predictions have been found in scaling
analyses of the field-tuned transitions in InO$_{x}$ \cite{Hebard}
and MoGe \cite{Yazdani}. However, the predicted universal critical
resistance is not observed in these experiments.

\begin{figure}[tb]
\centerline{\epsfig{file=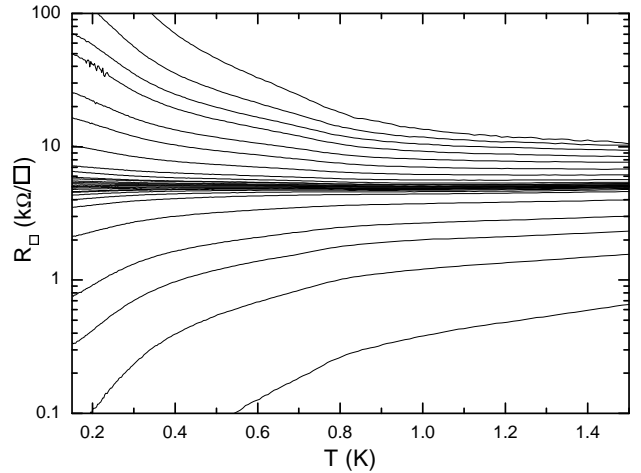,width=8.25cm}} \caption{Sheet
resistance versus temperature measured on one marginally
superconducting film at various magnetic field values. For curves
from bottom to top, the field increased from 0.05 T to 2.25 T. The
critical field was 0.66 T. In zero-field, $T_{c}$ $\sim$ 2 K.}
\label{Figure 1}
\end{figure}

In this Letter, we report studies of the field-tuned SI transition
in ultrathin Be films quench-condensed near 20 K. We describe
scaling analyses of the field-tuned transition based on the "dirty
boson" model. We have found that, for measurements performed in
the zero-current limit, the scaling exponent product $\nu z$ $=$
1.35 $\pm$ 0.10. This agrees well with results of the field-tuned
transition in relatively thicker InO$_{x}$ \cite{Hebard} and MoGe
\cite{Yazdani} films, but disagrees with a recent report
\cite{Markovic,Markovic2} of $\nu z$ $=$ 0.7 for the field-tuned
transition in ultrathin Bi/Ge films. For our Be films having a
robust superconducting phase, $R_{c}$ appears to be near
1.2$R_{Q}$ in the field-tuned transition. This agrees well with
results from the Bi/Ge films \cite{Markovic,Markovic2}. We have
also carried out for the first time studies of the field-tuned
transition in the presence of a dc bias current. The applied dc
bias current should exert a force of $\bf{j} \times \bf{B}$ on the
vortices in the direction perpendicular to both the applied
magnetic field $\bf{B}$ and the current density $\bf{j}$. We have
observed that the scaling exponent product $\nu z$ becomes 0.75
$\pm$ 0.10 in the presence of dc bias currents, suggesting that
the field-tuned transitions with and without dc bias current
belong to different universality classes.

In Fig. 1, we show, with zero bias current, the temperature
dependence of the sheet resistance, $R_{\Box}$, measured at
various field values for one of our Be films with the magnetic
field applied perpendicular to the plane of the film. These Be
films were quench-condensed onto bare glass substrates which were
held near 20 K during evaporations, under UHV conditions inside a
dilution refrigerator. This $\em{in}$ $\em{situ}$ progressive
evaporation setup allowed for systematic studies of the SI
transition as film thickness was varied. We deposited each set of
films in fine steps. We carefully monitored $R_{\Box}$ during each
evaporation step until a desirable value of $R_{\Box}$ was
reached. The films were very close to 10 {\AA} in thickness,
however, our quartz thickness monitor was not sensitive enough to
pick up the small thickness increments after each step. Film
resistance was measured in a standard four-terminal geometry using
a PAR-124A lock-in amplifier operating at 27 Hz. The ac probe
current was fixed at 1 nA. At the finite measuring temperatures in
the vicinity of the field-tuned transition, the {\em I-V}
characteristics in this low-current regime were linear. More
details regarding these Be films have been published elsewhere
\cite{Bielejec}. Quench-condensed Be films are chosen for this
study because such films were found to be nearly amorphous
\cite{Bielejec}. These Be films undergo a transition from
insulating to superconducting \cite{Bielejec} when the normal
state sheet resistance, $R_{N}$, is reduced below $\sim$ 10
k$\Omega/\Box$ with increasing film thickness. We have now studied
the field-tuned SI transition in several films of $R_{N}$ between
5.6 and 12 k$\Omega$/$\Box$. The superconducting transition
temperature, $T_{c}$, of these films varied between 0.5 to 4 K in
zero-field.

\begin{figure}[tb]
\centerline{\epsfig{file=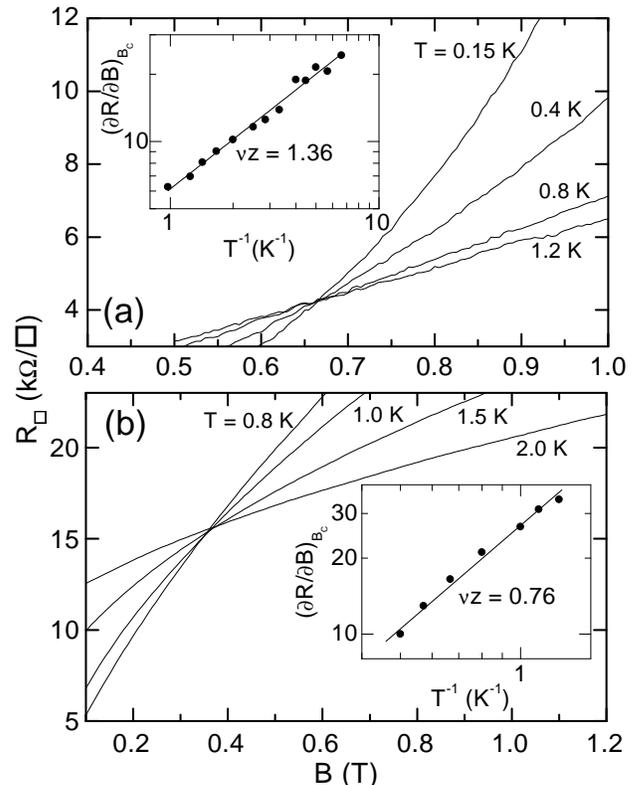,width=8.15cm}} \caption{Main
figures show sheet resistance vs. field at a number of
temperatures as labeled in the figures, with $I_{bias} =$ 0 in (a)
and $I_{bias} =$ 2.5 $\mu$A in (b). The crossing points identify
the critical fields and critical resistances. Insets show the
power-law relations between $\frac{\partial R}{\partial
B}\vert_{B_{c}}$ and $T^{-1}$ for the corresponding $I_{bias}$
values, which determine the values of $\nu z$.} \label{Figure 2}
\end{figure}

The film in Fig. 1, with $R_{N} =$ 10.7 k$\Omega$/$\Box$ at 15 K,
is considered marginally superconducting, as we will discuss
later. With increasing field, corresponding to curves from the
bottom to the top in Fig. 1, this film was driven from
superconducting to insulating, with a rather flat $R_{\Box}$ vs.
$T$ curve at a critical field of $B_{c} =$ 0.66 T. The main part
of Fig. 2(a) shows $R_{\Box}$ vs. $B$ at various temperatures for
the same film as shown in Fig. 1. In the vicinity of a quantum
critical point, the resistance of a two-dimensional system is
predicted to obey the scaling law in Eq. (1). Determining the
critical exponents involves plotting the $R_{\Box}$ vs. $B$ data
at various temperatures according to the scaling law. The good
crossing point, over one decade in temperature, in the $R_{\Box}$
vs. $B$ plot in the main part of Fig. 2(a) identifies $B_{c} =$
0.66 T and $R_{c} =$ 4.4 k$\Omega$/$\Box$ for this film. We have
used two methods to determine the scaling exponent $\nu z$. First,
we can find $\nu z$ by evaluating the derivative of the $R_{\Box}$
vs. $B$ curve at the critical field. In this case, we have the
following scaling equation:
\begin{equation}
\frac{ \partial R}{ \partial B} \vert_{B_{c}} \propto
R_{c}T^{-1/\nu z} f'(0).
\end{equation}
A plot of $\frac{ \partial R} {\partial B}\vert_{B_{c}}$ vs.
$T^{-1}$ on a log-log scale, shown in the inset to Fig. 2(a),
yields a straight line with a slope equal to 1/$\nu z$, from which
we determine $\nu z$ $=$ 1.36 $\pm$ 0.10. Alternatively, we can
plot $R$/$R_{c}$ vs. $\vert B - B_{c} \vert t$ and treat $t(T)$ as
an unknown variable. The values of $t(T)$ at various temperatures
are determined by obtaining the best collapse of the data.
Following this procedure presented in recent papers
\cite{Markovic,Markovic2} by N. Markovic $\em{et}$ $\em{al.}$, we
obtain the temperature dependence of $t(T)$ which we plot on a
log-log scale in the inset to Fig. 3(a). The collapse of the data
is shown in the main part of Fig. 3(a) in a $R$/$R_{c}$ vs. $\vert
B - B_{c}\vert t$ plot. Figure 3(a) shows good collapse of the
data over three orders of magnitude in $\vert B - B_{c}\vert t$
and two orders of magnitude in $R$/$R_{c}$. The straight line in
the inset to Fig. 3(a) shows a power-law fit, as expected by the
scaling function, Eq. (1). The slope of the line in the inset
gives -1/$\nu z$, from which we find $\nu z$  $=$ 1.34 $\pm$ 0.10.
The exponents obtained from the above two methods agree with each
other, showing the consistency of the scaling analysis. The
scaling exponents obtained in our Be films in the zero-bias limit
appear to be in very good agreement with the predictions of
theories based on the "dirty boson" model \cite{Fisher2,Fisher3}.
Our result is also in very good agreement with renormalization
group theories \cite{Zhang,Singh} and Monte Carlo \cite{Cha}
simulations.

\begin{figure}[ht]
\centerline{\epsfig{file=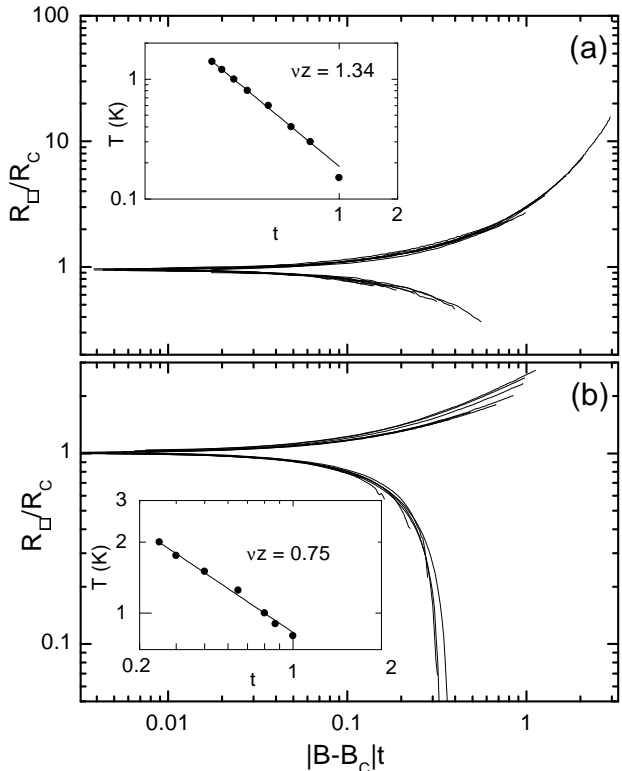,width=8.15cm}} \caption{Main
figures show the scaling plots of $R$/$R_{c}$ vs. $\vert B -
B_{c}\vert t$, with $I_{bias} =$ 0 in (a) and $I_{bias} =$ 2.5
$\mu$A in (b).  Insets show the power-law relations between the
parameter $t$ and the temperature, which determine the values of
$\nu z$.} \label{Figure 3}
\end{figure}

Another important prediction of the bosonic model is the universal
critical resistance $R_{c}$ $=$ $R_{Q} = h/4e^{2}$ $\approx$ 6.5
k$\Omega$/$\Box$ at the quantum critical point. This remains a
controversial issue since only in the Bi/Ge system
\cite{Markovic,Markovic2} has a critical sheet resistance close to
the predicted value of $R_{Q}$ been observed. In fact, careful
investigations in the Bi/Ge system have revealed that $R_{c}$ is
1.1$R_{Q}$ to 1.2$R_{Q}$ and decreases in a narrow range as
$R_{N}$ is reduced with increasing film thickness.  It was
suggested \cite{Markovic} that the slightly larger $R_{c}$ than
$R_{Q}$ was consistent with calculations in two-dimensions for a
bosonic model including Coulomb interactions \cite{Herbut}, which
predicted $R_{c}$ $\sim$ 1.4$R_{Q}$, as well as Monte Carlo
simulations in the (2+1)-dimensional $XY$ model without disorder
\cite{Cha2}, which found $R_{c}$ $=$ 1.2$R_{Q}$. The small
variation of $R_{c}$ with $R_{N}$ could be explained
\cite{Markovic2} as a finite-temperature effect, since strictly
speaking the critical resistance is predicted to be universal only
at $T =$ 0. In other systems, such as MoGe \cite{Yazdani} films,
$R_{c}$ varies in a much wider range among films of varying
$R_{N}$. In order to explain the low $R_{c}$ of their MoGe films,
Yazdani and Kapitulnik proposed \cite{Yazdani} a two-channel model
in which the conductance due to normal electrons add to the
conductance due to the Cooper pairs. If the unpaired electrons are
localized, this model collapses to a single channel bosonic model
with $R_{c}$ $=$ $R_{Q}$. Otherwise, unpaired electrons add to the
conduction at the transition. It has been noted \cite{Markovic2}
that the resulting film resistance can be either larger or smaller
than $R_{Q}$ since the magnetoresistance contribution
\cite{Bergmann} from the unpaired electrons can be either
positive, if the spin-orbit coupling is strong, or negative, if
the spin-orbit coupling is very weak.

Our Be films showed a zero-field SI transition as $R_{N}$ was
reduced below $\sim$ 10 k$\Omega$/$\Box$ with increasing film
thickness \cite{Bielejec}. Films of $R_{N}$ near 11
k$\Omega$/$\Box$ are considered marginally superconducting since
they have near-zero critical field values. Further increasing film
thickness, the critical field, $B_{c}$, increases with decreasing
$R_{N}$, as shown in Fig. 4(a). In Fig. 4(b), we plot how $R_{c}$
in the field-tuned transition varies with $R_{N}$. For marginally
superconducting films of $R_{N}$ between 9 to 12 k$\Omega$/$\Box$,
$R_{c}$ was significantly smaller than $R_{Q}$. Nevertheless, we
see clearly by comparing Fig. 4(a) with Fig. 4(b) that, as $R_{N}$
was reduced by increasing film thickness, films with robust
critical field values have a $R_{c}$ of 7.1 $\sim$ 8.0
k$\Omega$/$\Box$ in the field-tuned SI transition, which is about
1.2$R_{Q}$. This result is in excellent agreement with results
from the Bi/Ge system \cite{Markovic2}. However, our results
disagree with the suggestion \cite{Markovic2} that the
magnetoresistance of the unpaired electrons caused the discrepancy
between $R_{c}$ and $R_{Q}$. Since Be has the weakest spin-orbit
coupling among metals, the magnetoresistance should be negative,
leading to $R_{c}$ $<$ $R_{Q}$ following the two-channel model. We
note that we have shown in Fig. 1, Fig. 2(a), and Fig. 3(a) data
measured on a marginally superconducting film of $R_{N} =$ 10.7
k$\Omega$/$\Box$ at 15 K. Although the critical resistance of this
film, $R_{c}$ $\sim$ 4.4 k$\Omega$/$\Box$, is significantly
smaller than $R_{Q}$, we have found that, for all the marginally
and robust superconducting films of $R_{N}$ ranging from 5.6 to 12
k$\Omega$/$\Box$, the critical exponents are the same with $\nu z$
$=$ 1.35 $\pm$ 0.10 in the zero-current limit.

\begin{figure}[t]
\centerline{\epsfig{file=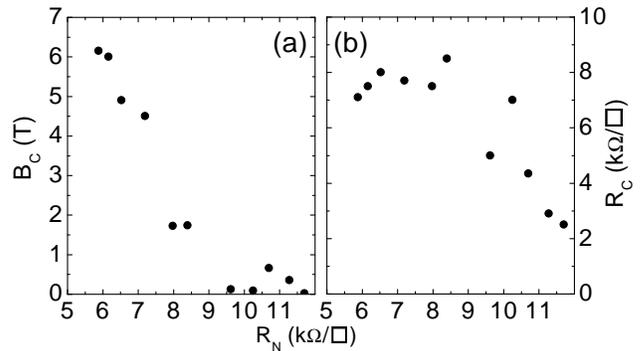,width=8.25cm}}
\caption{Figures show (a) the critical field, $B_{c}$, and (b) the
critical resistance, $R_{c}$, as functions of the normal-state
sheet resistance, $R_{N}$, measured at 15 K. } \label{Figure 4}
\end{figure}

Below, we describe results of the field-tuned SI transition in the
presence of a dc bias current, $I_{bias}$. For such studies,
$I_{bias}$ was varied between 125 nA and 2.5 $\mu$A and kept below
the zero-field critical current ($\sim$ 15 $\mu$A). At each fixed
$I_{bias}$, we used a magnetic field to tune a film from
superconducting to insulating.  We believe that joule heating was
insignificant in our experiments based on the following arguments.
First, we have performed extensive {\em I-V} measurements over the
entire temperature and magnetic field range of our experiments,
with the time span for the {\em I-V} sweeps ranging from ten
minutes to one hour. The {\em I-V} curves were completely
reproducible without any observable hysteresis and independent of
the sweep rate. Secondly, we can estimate the temperature
increase, $\Delta T$, on the Be films due to joule heating. The
thin Be films were deposited on glass microslide substrates of
thickness 0.23 mm, which was attached to a copper sample holder by
a very thin layer of grease. The glass substrate was the dominant
source of heat resistance, with a thermal conductivity of $\sim$
0.0003 W/Km at 100 mK \cite{White}. For a typical film square of
size 3$\times$3 mm$^{2}$ and resistance 10 k$\Omega$, joule
heating for $I_{bias}=$ 125 nA is about 0.15 nW, resulting in
$\Delta T \sim$ 0.012 mK at 100 mK. For $I_{bias}=$ 2.5 $\mu$A,
$\Delta T \sim$ 5.0 mK at 100 mK. We note that the heating power
at 2.5 $\mu$A was 400 times larger than the heating power at 125
nA. If joule heating were significant, the data obtained with
$I_{bias}=$ 2.5 $\mu$A should show a flattening of the data in the
low temperature region when compared to the data obtained with
$I_{bias}=$ 125 nA. The fact that the scaling results at 125 nA
and 2.5 $\mu$A agree well suggests that heating was insignificant.
This also argues against the existence of significant electron
heating decoupled from the lattice.

We plot in Fig. 2(b) and Fig. 3(b), for $I_{bias} =$ 2.5 $\mu$A,
the results of scaling analyses based on the two previously
described methods. In Table I, we list the parameters from scaling
analyses of the field-tuned transition at various $I_{bias}$.
Results from the data collapsing method are presented for
$I_{bias}$ values of 250 nA and 2.5 $\mu$A, for which the amount
of data taken was adequate for such analyses. It appears that $\nu
z$ $\sim$ 0.75 $\pm$ 0.10, showing no systemic change with
$I_{bias}$ in the range we have studied. Nevertheless, it is
significantly smaller than the $\nu z$ found in the zero-current
limit. We can only speculate that the bias current could lead to a
symmetry-breaking, resulting in different critical behavior at the
transition. We note that experiments need to be carried out in
other systems in order to find out whether this result is
universal. In addition, experiments at low $I_{bias}$ values
should be carried out to determine whether $\nu z$ changes
abruptly or gradually as $I_{bias}$ is increased from zero. Such
experiments can probe the threshold $I_{bias}$ for the change in
the scaling exponents, and have the potential of revealing whether
the new scaling exponents are produced by certain nonlinear
effects in the vortices under a bias current. On the other hand,
we need to discuss other possible origins that might have led to
the new scaling exponents. For example, transport in marginally
superconducting films, or, in the presence of a dc bias, could be
dominated by narrow superconducting filaments, leading to a change
in the dimensionality of the system. While it is difficult to
determine to what extent film inhomogeneity affects our
experiments, we would like to comment on the critical currents
measured in our films. For the film shown in Fig. 2(b) and Fig.
3(b), the critical current in zero field was about
1.5$\times$10$^{-5}$ A. This film was about 10 {\AA} in thickness
and 3 mm in width. Thus, the critical current density would be
about 5$\times$10$^{6}$ A/m$^{2}$ if the current were uniformly
distributed in the bulk of the film. On the other hand, if we
assume that the current runs through a few filaments of 100 {\AA}
in total width, the critical current density would be
1.5$\times$10$^{12} $A/m$^{2}$. The critical currents of amorphous
Be films have been measured by other groups, for example by
Okamoto {\em et al.} \cite{Okamoto}. Their Be films had a
resistivity of 3.6$\times$10$^{-6}$ $\Omega$m. This value gives
rise to $R_{\Box} \sim$ 3.6 k$\Omega$/${\Box}$ for a film of
thickness same as ours ($\approx$ 10 {\AA}), meaning that their
films were roughly 3 times less resistive than ours. They found
that their films had critical current density of about
1$\times$10$^{8}$ A/m$^{2}$. Compared to their critical current
density, the assumption in our films of conduction dominated by a
few narrow filaments leads to an unreasonably large critical
current density. In addition, we can compare our marginally
superconducting films with robust superconducting films which are
unlikely to be dominated by narrow filaments. As we have
discussed, with $I_{bias} =$ 0, these two types of films showed
identical scaling exponents $\nu z$. It is possible that film
inhomogeneity has lead to a depressed $R_{c}$ in marginally
superconducting films, but has not changed the effective
dimensionality of the system. Therefore, the change in the scaling
exponents with a dc bias is unlikely due to a dimensional
cross-over as a result of filamentary superconductivity.

In conclusion, the scaling exponents found in our Be films in the
zero-biased, field-tuned SI transition agree very well with
results from InO$_{x}$ \cite{Hebard} and MoGe \cite{Yazdani}
films, but disagrees with recent results \cite{Markovic,Markovic2}
from quench-condensed Bi/Ge films. For Be films having a robust
superconducting phase, the critical sheet resistance in the
field-tuned SI transition was about 1.2$R_{Q}$. Our field-tuned
transition in the presence of a dc current reveals a set of new
scaling exponents, suggesting that the field-tuned transitions
with and without dc bias belong to different universality classes.
We gratefully acknowledge S. Teitel and Y. Shapir for numerous
useful discussions.

\begin{table}[t]
\caption{Bias currents and derived parameters} \label{Table 2}
\begin{tabular}{|cr|c|c|c|c|c|}
$I_{bias}$ (nA) & & $B_{c}$ (T) & $R_{c}$ (k$\Omega$/$\Box$) &
$R_{N}$ (k$\Omega$/$\Box$) & $\nu z$ \footnote{$\nu z$ obtained by
the $\frac{\partial R}{\partial B}\vert_{B_{c}}$ method}  & $\nu
z$ \footnote{$\nu z$ obtained by the data collapse method}
\\
\hline

125  & &   0.33   &  15.5 & 9.36 & 0.81 & -- \\

250  & &   0.35   &  16.1 & 9.36 & 0.76 & 0.73 \\

1000  & &   0.36   &  15.5 & 9.36 & 0.73 & -- \\

2500  & &   0.38   &  15.0 & 9.36 & 0.77 & 0.75 \\

\end{tabular}

\footnotemark[1]{$\nu z$ obtained by the $\frac{\partial
R}{\partial B}\vert_{B_{c}}$ method}

\footnotemark[2]{$\nu z$ obtained by the data collapse method}
\end{table}
%
%
%REFERENCES HERE
%


%Here is an example of the general form of a table:


\begin{references}

\bibitem{Sondhi} S. L. Sondhi {\em et al.}, Rev. Mod. Phys. {\bf69}, 315 (1997).

\bibitem{Fisher} M. P. A. Fisher {\em et al.}, Phys. Rev. B {\bf40}, 546 (1989).

\bibitem{Fisher2} M. P. A. Fisher, Phys. Rev. Lett. {\bf65}, 923 (1990).

\bibitem{Fisher3} M. P. A. Fisher, G. Grinstein, and S. M. Girvin, Phys. Rev.
Lett. {\bf64}, 587 (1990).

\bibitem{Cha} M.-C. Cha {\em et al.}, Phys. Rev. B {\bf44}, 6883 (1991)

\bibitem{Valles} J. M. Valles, Jr., R. C. Dynes, and J. P. Garno, Phys. Rev.
Lett. {\bf69}, 3567 (1992); S-Y. Hsu, J. A. Chervenak, and J. M.
Valles, Jr., {\em ibid.} {\bf75}, 132 (1995).

\bibitem{Hebard} A. F. Hebard and M. A. Paalanen, Phys. Rev. Lett. {\bf65}, 927
(1990); M. A. Paalanen, A. F. Hebard, and R.R. Ruel, {\em ibid}
{\bf69}, 1604 (1992).

\bibitem{Yazdani} Ali Yazdani and Aharon Kapitulnik, Phys. Rev. Lett. {\bf74}, 3037
(1995); D. Ephron, A. Yazdani, A. Kapitulnik, and M. R. Beasley,
{\em ibid.} {\bf76}, 1529 (1996).

\bibitem{Markovic} N. Markovic, C. Christiansen, and A. M. Goldman, Phys. Rev.
Lett. {\bf81}, 5217 (1998).

\bibitem{Markovic2} N. Markovic {\em et al.}, Phys. rev. B {\bf60}, 4320 (1999).

\bibitem{Bielejec} E. Bielejec, J. Ruan, and Wenhao Wu, Phys. Rev. B {\bf63}, 100502
(2001); E. Bielejec, J. Ruan, and Wenhao Wu, Phys. Rev. Lett.
{\bf87}, 036801 (2001).

\bibitem{Zhang} L. Zhang and M. Ma, Phys. Rev. B {\bf45}, 4855 (1992).

\bibitem{Singh} K. G. Singh and D. S. Rokhsar, Phys. Rev. B {\bf46}, 3002 (1992).

\bibitem{Herbut} Igor F. Herbut, Phys. Rev. Lett. {\bf81}, 3916 (1998).

\bibitem{Cha2} M. -C. Cha and S. M. Girvin, Phys. Rev. B {\bf49}, 9794 (1994).

\bibitem{Bergmann} G. Bergmann, Phys. Rep, {\bf107}, 1 (1984).

\bibitem{White} G. K. White, {\em Experimental Techniques in Low
Temperature Physics}, 3rd ed., (Clarendon Press, Oxford, 1979), p.
320.

\bibitem{Okamoto} M. Okamoto, K. Takei, and S. Kubo, J. Appl. Phys. {\bf62}, 212 (1987)

\end{references}
\end{document}